\documentstyle[epsf,twoside,fleqn,espcrc2]{article}

\newcommand{\AmS}{{\protect\the\textfont2
  A\kern-.1667em\lower.5ex\hbox{M}\kern-.125emS}}

\title{Non-perturbative Renormalization Constants using Ward Identities\thanks{
        This work was supported by the DoE Grand Challenges award at the ACL at Los Alamos}
        }
\author{Tanmoy Bhattacharya\address{MS B-285, 
            Los Alamos National Lab, Los Alamos, New Mexico 87545, USA}, 
	Shailesh Chandrasekharan\address{Department of Physics, 
		Duke University, Durham,
        	North Carolina 27705, USA}, 
	Rajan Gupta${}^{\rm a}$, Weonjong Lee${}^{\rm a}$, 
	Stephen Sharpe\address{Physics Department, University of Washington,
        	Seattle, Washington 98195, USA}
       }

\begin{document}

\begin{abstract}
We extend the application of vector and axial Ward identities to
calculate $b_A$, $b_P$ and $b_T$, coefficients that
give the mass dependence of the renormalization constants of 
the corresponding bilinear operators in the quenched theory. 
The extension relies on using operators with non-degenerate quark masses. 
It allows a complete determination of the $O(a)$ improvement coefficients
for bilinears in the quenched approximation using Ward Identities alone.
Only the scale dependent normalization constants $Z_P^0$ (or $Z_S^0$)
and $Z_T$ are undetermined.
We present results of a pilot numerical study using hadronic correlators.
\end{abstract}

\maketitle


To remove errors of $O(a)$ from physical matrix elements, 
one must improve both the action and the operators~\cite{symanzik}. 
The former requires the addition 
of the Sheikholeslami-Wohlert term~\cite{sw-1},
\begin{equation}
	S_{SW} = - a^5 c_{SW}\ \sum_{x} \bar{\psi}(x) \frac{i}{4}
	\sigma_{\mu\nu} F_{\mu\nu}(x) \psi(x) \,.
\end{equation}
Improvement of flavor off-diagonal bilinears requires \cite{alpha-0}
both the addition of extra operators,
\begin{eqnarray}
(A_I)_{\mu}    & \equiv & A_{\mu} + a c_A \partial_\mu P \\
(V_I)_{\mu}    & \equiv & V_{\mu} + a c_V \partial_\nu T_{\mu\nu} \\
(T_I)_{\mu\nu} & \equiv & T_{\mu\nu} +
                a c_T ( \partial_\mu V_\nu - \partial_\nu V_\mu) \,,
\end{eqnarray}
and the introduction of a mass dependence,
\begin{eqnarray}
(X_R)^{(ij)}    & \equiv & Z_X^0(1+b_X am^{ij}) (X_I)^{(ij)} \,.
\label{def_XA} 
\end{eqnarray}
Here $X= A, V, P, S, T$, $Z_X^0$ are the renormalization constants 
in the chiral limit, and  $m_{ij} \equiv ( m_i + m_j)/2$ 
are the bare quark masses defined using the axial Ward Identity
(WI), Eq.~(\ref{cA}).
The bare unimproved bilinears are
$ A^{(ij)}_{\mu}    \equiv \bar{\psi^i} \gamma_{\mu} \gamma_{5} \psi^j $, 
etc.
The task is to determine the coefficients $c_{SW}$, $Z_X^0$'s, 
$c_X$'s, and $b_X$'s non-perturbatively.

Previous calculations have shown how 
$Z_V^0$, $Z_A^0$ and $Z_P^0/Z_S^0$ \cite{rome-0}, 
$c_{SW}$, $c_A$ and $b_V$ \cite{alpha-1,alpha-2,alpha-3}, 
$c_V$ \cite{alpha-4}, $c_T$ \cite{rome-1},
and $b_P - b_A$ and $b_S$ \cite{rome-3}
can be determined non-perturbatively using axial and vector WI. 
We discuss here an extension that yields $b_A$, $b_P$, and $b_T$.
The two remaining constants $Z_P^0$ (or $Z_S^0$) and $Z_T^0$ are scale
and scheme dependent, and so cannot be determined using WI.  Note that
the relations we derive do not extend directly to the unquenched
theory, which requires additional improvement constants and a more
complicated set of conditions~\cite{WIunquenched}.

%
%
We begin by recalling the ALPHA method for determining
$c_{SW}$ and $c_A$ \cite{alpha-2}. The bare WI mass
\begin{equation}
2 m_{ij} =  \frac{ \sum_{\vec{x}} \langle 
 \partial_\mu [A_\mu + 
  a c_A \partial_\mu P]^{(ij)}(\vec{x},t) J^{(ji)}(0) \rangle} 
 {\sum_{\vec{x}} \langle P^{(ij)}(\vec{x},t) J^{(ji)}(0) \rangle} 
\label{cA} 
\end{equation}
should be independent of $t$, up to corrections of $O(a^2)$,
since it is proportional to the average renormalized quark mass
\begin{equation}
\frac{m^R_i+m^R_j}{2} =  m_{ij} \frac{Z_A^0(1+b_A a m_{ij})} 
{Z_P^0(1+b_P am_{ij})}  \,.
\label{m(t)}
\end{equation}
This is achieved by simultaneously tuning $ c_{SW} $ and $c_A$.  Our
approach differs from the Schr\"odinger functional method of
Ref.~\cite{alpha-2} in that we use standard 2-point correlation
functions.  Consistency of these estimates are checked by varying the
initial state using $J = P$ or $A_4$ and with different types of
sources for the quark propagators (Wuppertal smearing, Wall, point).
In the following we use the abbreviation $m_i = m_{ii}$, where
$m_{ii}$ refers to two degenerate flavors.

%
%
With $c_{SW}$ fixed,
{\bf $ Z_V^0$} and {\bf $b_V$} are obtained using charge conservation.
We use the forward matrix elements of $(V_I)_4$ between pseudoscalars,
\begin{eqnarray}
\lefteqn{\frac{1}{ Z_V^0 (1+b_V am_2) } = } \nonumber \\
& & \mbox{} \frac{ \sum_{\vec{x}, \vec{y}}
  \langle P^{(12)}(\vec{x},\tau) (V_I)_4^{(22)}(\vec{y},t) J^{(21)}(0) \rangle }
  { \langle \sum_{ \vec{x}} P^{(12)}(\vec{x},\tau) J^{(21)}(0) \rangle } \,.
\label{ZV}
\end{eqnarray}
%
with $ \tau > t > 0 $ and $ J = P $ or $A_4$.
Note that the $c_V$ term in $V_I$ does not contribute.

%
%
Next consider the generic axial WI
\begin{equation}
\left\langle 
  \delta {\cal S}^{(12)} {\cal O}_R^{(23)}(y) J^{(31)}(z) \right\rangle
  = \left\langle \delta {\cal O}_R^{(13)}(y) \  
J^{(31)}(z)\right\rangle 
\label{AWIdef}
\end{equation}
where $ {\cal O}^{23} = {\bar\psi}^{(2)} \Gamma \psi^{(3)} $, 
$ \delta {\cal O}^{13} = {\bar\psi}^{(1)} \gamma_5 \Gamma \psi^{(3)} $, 
and 
\begin{equation}
 \delta {\cal S}^{(12)} = \int_{\cal V} 
    \left[ (m^R_1+m^R_2) (P_R)^{(12)}\! - \!
\partial_\mu (A_R)^{(12)}_\mu \right] \hskip -2pt
\label{AWI}
\end{equation}
%
%
This results from a chiral rotation on flavors $1,2$ in the 4-volume
$\cal V$, with $ y \in {\cal V} $ and $ z \not\in {\cal V} $.
By enforcing these identities in the chiral limit one can determine
$c_V$ and $c_T$ \cite{alpha-0,alpha-4,rome-1}, as shown below.

Away from the chiral limit, operators $P$ and ${\cal O}$ in the
product $\int_{\cal V} (m^R_1+m^R_2) (P_R)^{(12)}(x) {\cal
O}_R^{(23)}(y)$ need off-shell improvement. This requires the
addition of a contact term, of unknown normalization, having the same
form as the RHS of (\ref{AWIdef})~\cite{rome-1}.
Our new observation is that the contact term is proportional to 
$m_1+m_2$ and so can be removed
by extrapolating $m_1$ and $m_2$ to zero.
This leaves the freedom to examine the dependence on $m_3$,
and from this one can determine certain combinations of the $b_X$.
In the following, the extrapolation to $m_1=m_2=0$ is implicit\footnote{%
In practice, we keep the $\int (m_1+m_2) (P_R)^{(12)}$ term in Eq. (\ref{AWI})
prior to extrapolation, since it improves the signal.}.

%
%
As a first application of this method we show how to obtain $b_A$,
as well as $c_V$, using the AWI
\begin{eqnarray}
\lefteqn{r_1 \equiv  \frac{ Z_A^0 (1+b_A am_3/2) } 
{ Z_A^0 \cdot Z_V^0 (1+ b_V am_3/2) } }
\nonumber \\
 &=&  \frac{ \sum_{\vec{y}}
	\langle \delta {\cal S}^{(12)} \ 
	(V_I)_4^{(23)}(\vec{y},y_4) \  J^{(31)}(0) \rangle }
{ \sum_{\vec{y}} \langle (A_I)_4^{(13)}(\vec{y},y_4) \ J^{(31)}(0) \rangle } 
\label{cV1} \\
&=& \frac{ \sum_{\vec{y}} e^{i\vec{p} \cdot \vec{y} } 
        \langle \delta {\cal S}^{(12)} 
	(V_I)_i^{(23)}(\vec{y},y_4)
	A_i^{(31)}(0) \rangle }
        { \sum_{\vec{y}} e^{i\vec{p} \cdot \vec{y} }
		\langle (A_I)_i^{(13)}(\vec{y},y_4) 
		A_i^{(31)}(0) \rangle } 
\label{cV2} 
%
\end{eqnarray}  
where $J = P$ or $A_4$.  Eq.~(\ref{cV1}) is independent of $c_V$ at
$\vec p = 0$, and its $m_3$ dependence gives $b_A-b_V$. The intercept
provides a second determination of $Z_V^0$.  Eq.~(\ref{cV2}) is used
to determine $c_V$.

%
%
Given $c_V$, an alternate determination of $b_A-b_V$ 
is obtained from
\begin{eqnarray}
\lefteqn{r_2 \equiv 
\frac{ Z_V^0 (1+b_V am_3/2) } { Z_A^0 \cdot Z_A^0 (1+ b_A am_3/2) }}
\\
 &=&  \frac	{ \sum_{\vec{y}} e^{i\vec{p} \cdot \vec{y} }
	\langle \delta {\cal S}^{(12)} \ (A_I)_i^{(23)}(\vec{y},y_4) \  V_i^{(31)}(0) \rangle }
	{ \sum_{\vec{y}}  e^{i\vec{p} \cdot \vec{y} }
	\langle (V_I)_i^{(13)}(\vec{y},y_4) \ V_i^{(31)}(0) \rangle } \nonumber 
\label{ZAZV-1}
\end{eqnarray}
This also yields $Z_A^0$.
The same information can be obtained from the combinations
\begin{eqnarray}
	& & \frac{1}{ \sqrt{ r_1 \cdot r_2} } = Z_A^0 \,,
\label{ZAZV-2}
\\
	& & \sqrt{ \frac{r_1} {r_2} } =
	\frac{Z_A^0}{Z_V^0} (1+ (b_A - b_V) am_3/2) \,.
\label{ZAZV-3}
\end{eqnarray}
%
%
Similarly, we determine $b_P - b_S$ and $ Z_P^0/Z_S^0 $ using
\begin{eqnarray}
\lefteqn{
\frac{ Z_P^0 (1+b_P am_3/2) } { Z_A^0 \cdot Z_S^0 (1+ b_S am_3/2) }}
\nonumber \\
&& \hskip -0.2 in =  \frac{ \sum_{\vec{y}} e^{i\vec{p} \cdot \vec{y} }
\langle \delta {\cal S}^{(12)} \ S^{(23)}(\vec{y},y_4) \ J^{(31)}(0) \rangle }
	{ \sum_{\vec{y}}  e^{i\vec{p} \cdot \vec{y} }
	\langle P^{(13)}(\vec{y},y_4) \ J^{(31)}(0) \rangle } \,.
	\label{ZPZS-1}
\end{eqnarray}

%
To get $ c_T $ we use the WI with ${\cal O} = T_{ij}$ 
\begin{eqnarray}
\lefteqn{1 + a c_T \frac{ \sum_{\vec{y}} \langle
 [- \partial_4 V_k ]^{(13)}(\vec{y},y_4) T_{k4}^{(31)}(0) \rangle }
{ \sum_{\vec{y}} \langle T_{k4}^{(13)} (\vec{y},y_4) T_{k4}^{(31)}(0) 
\rangle } }
\nonumber \\
& & \hskip -0.2 in = Z_A^0 \frac { \sum_{\vec{y}} 
        \langle \delta {\cal S}^{(12)} \ (T_I)_{ij}^{(23)}(\vec{y},y_4) 
	\ T_{k4}^{(31)}(0) \rangle }
        { \sum_{\vec{y}}  
        \langle T_{k4}^{(13)} (\vec{y},y_4) \ T_{k4}^{(31)} (0) \rangle } \,.
        \label{cT-1}
\end{eqnarray}
At $\vec p = 0$, $ (T_I)_{ij}$ has no contribution
from the $ c_T $  term, so the only $c_T$ dependence is on the LHS.

%
%
The previous method fails for $b_T$ since both sides of (\ref{AWIdef})
have the same dependence on $b_T$ if $m_1=m_2=0$.
The cure is to consider three non-degenerate masses.
The contact term required to improve the LHS of (\ref{AWIdef}) is
proportional to $m_1+m_2$, while the relative dependence of the
two sides on $b_T$ is proportional to $m_1-m_2$.
Thus the two terms can be separated and $b_T$ determined.
More details will be given in \cite{WIunquenched}.
We have not yet implemented this proposal.

%
%
Thus far we do not have a separate determination of $b_P$ or $b_S$.
This can be accomplished with the method of Ref.~\cite{rome-3}, which uses
2-point correlation functions for non-degenerate masses. In addition to 
using that method we present a variant which avoids the need to study
quantities as a function of the underlying hopping parameter $\kappa$.
%
%
We first note that if we use Eq.~(\ref{m(t)}) and
enforce $(2 m_1)^R + (2 m_2)^R = 2(m_1+m_2)^R$, we find
\begin{equation}
b_P - b_A = -
	\frac{ 4m_{12} - 2 [ m_{11} + m_{22} ]}
	{  a [ m_{11} - m_{22} ]^2 }  \,.
\label{bP-bA}
\end{equation}
We next make use of the vector two-point WI
\begin{eqnarray*}
\Delta m_{12} &\equiv&
\frac{ \sum_{\vec{x} }  e^{ i\vec{p} \cdot \vec{x} }
	\langle \partial_\mu {V_I}_\mu^{(12)} (\vec{x},t) J^{(21)}(0) \rangle }
        { \sum_{\vec{x} }  e^{ i\vec{p} \cdot \vec{x} }
	\langle S^{(12)}(\vec{x},t) J^{(21)}(0) \rangle }
\label{VS1}
\\
m^R_1 - m^R_2 &=& 
\frac	{ Z_V^0 [1 + b_V a (m_1+m_2)/2] } 
	{ Z_S^0 [1 + b_S a (m_1+m_2)/2] } \Delta m_{12} 
\label{VS2}
\end{eqnarray*}
where the source $J$ is either $ J^{(21)}= S^{(21)}$ or
$\sum_{\vec{z}} P^{(23)}(\vec{z},z_4) P^{(31)}(0) $ for $ 0 < t < z_4$.
Enforcing $2(m_1^R-m_2^R) = (2 m_1)^R-(2m_2)^R$, we find
\begin{eqnarray}
\frac{b_S - b_V}{2} &+&  (b_P - b_A ) \nonumber \\
       &=&  \frac{ \Delta m_{12} - R_Z [ m_{11} - m_{22} ] }
	{ a R_Z [ m_{11}^2 - m_{22}^2 ]}       \,, 
\label{bS-bV}
\end{eqnarray}
where $R_Z \equiv {Z_S^0}Z_A^0/({Z_P^0}Z_V^0)$.
Since $ b_A $ and $b_V$ are already known,
$b_P$ and  $b_S$ are given by Eqs.~(\ref{bP-bA}) and (\ref{bS-bV}).

The above discussion shows that {\it in principle} one can determine
all the constants, except $Z_P^0$ (or $Z_S^0$) and $Z_T^0$, in the
quenched theory using Ward identities.  The results of an exploratory
study are summarized in Table~\ref{t:Zresults}. These were obtained on
83 lattices of size $16^3 \times 48$ at $\beta=6.0$. Since the action
is only tree-level tadpole improved ($c_{SW} = 1.4755$), the results
do not represent full $O(a)$ improvement, but they indicate
the efficacy of the method. More details will appear in \cite{WIunquenched}.

%
\begin{table}[t]
\begin{center}
\begin{tabular}{| c | c | l |}
\hline
Eq.\# 	&	observable 	&	intercept	\\ 
\hline
(\ref{cA})	&	$ c_A $ 	&	$-0.016(11)$\\ 
\hline
(\ref{ZV})	&	$ Z^0_V $ 	& 	$+0.746(1) $\\ 
\hline
(\ref{ZV})	&	$ b_V $ 	&	$+1.55(2)$\\ 
\hline
(\ref{cV1}) 	&	$ Z^0_V $ 	&       $+0.752(7)$\\ 
\hline
(\ref{cV1})	&	$ b_A - b_V $ 	&	$+0.34(21)$\\ 
\hline
(\ref{cV2})	&	$ c_V $  	&	$+0.46(29)$\\ 
\hline
(\ref{ZAZV-1})	& $ Z^0_V/(Z^0_A)^2 $	&	$+1.32(12)$\\ 
\hline
(\ref{ZAZV-1})	&	$ b_A - b_V $	&	$+1.8(1.1) $\\ 
\hline
(\ref{ZAZV-2})	&	$ Z^0_A $	&	$ +0.78(2)$\\ 
\hline
(\ref{ZAZV-3})	&	$ Z^0_A/Z^0_V $	&	$ +1.00(5) $\\ 
\hline
(\ref{ZAZV-3})	&	$ b_A - b_V $	&	$ +1.2(8)$\\ 
\hline
(\ref{ZPZS-1})	& $Z^0_A Z^0_S/Z^0_P$	&	$+0.96(1) $\\ 
\hline
(\ref{ZPZS-1})	& 	$ b_P - b_S $ 	&	$-0.08(9) $\\ 
\hline
(\ref{cT-1})	& 	$ c_T $		&	$-0.14(7) $\\ 
\hline
\cite{rome-3}	& $Z^0_A Z^0_S/Z^0_P$	&	$+0.96(1) $\\ 
\hline
\cite{rome-3}	& $b_A-b_P+b_S/2$ 	&	$+0.49(1) $\\ 
\hline
(\ref{bP-bA})	& 	$b_P - b_A$ 	&	$+0.1(4)$\\ 
\hline
(\ref{bS-bV})	& $b_S-b_V-2(b_P-b_A)$	&	$-0.5(5)$\\
\hline
\end{tabular}
\end{center}
\caption{Constants extracted from the different WI.}
\label{t:Zresults}
\vspace{-0.2in}
\end{table}

We draw two preliminary conclusions.  First, even though we have found
channels in which the statistical and systematic errors on the
determination of $b_A - b_V$ and $b_P - b_S$ are fairly small, the
magnitude of these differences are still comparable to their error.
This rough equality between all the $b_X$ is consistent with
perturbative results~\cite{LuscherbSbm}.  Second, the determination of
$c_V$ has a large uncertainty, which accounts for a substantial
fraction of the errors in $Z_A^0$, $Z_P^0 / Z_S^0$, and $c_T$.  For
$c_A$ and $c_V$, the
Schr\"odinger functional method~\cite{alpha-1,alpha-2,alpha-3,alpha-4}
gives results with much smaller errors, and may prove to be the
method of choice.


%

\end{document}